# Topologically different spin disorder phases of the $J_1-J_2$ Heisenberg model on the honeycomb lattice

Jing Liu [1,2], Ya-Min Quan [1], H. Q. Lin [3,4] and Liang-Jian Zou [1,2,#]

1 *Key Laboratory of Materials Physics, Institute of Solid State Physics, Chinese Academy of Sciences, P. O. Box 1129, Hefei 230031, China*

2 *University of Science and Technology of China, Hefei 230026, China*

3 *Beijing Computational Science Research Center, Beijing 100193, China*

4 *Department of Physics, Beijing Normal University, Beijing, 100875, China*

## Abstract:

Searching for spin liquids on the honeycomb $J_1$-$J_2$ Heisenberg model has been attracting great attention in the past decade. In this Paper we investigate the topological properties of the $J_1$-$J_2$ Heisenberg model by introducing nearest-neighbour and next-nearest-neighbour bond parameters. We find that there exist two topologically different phases in the spin disordered regime $0.2<J_2/J_1<0.5$: for $J_2/J_1<0.32$, the system is a zero-flux spin liquid which is topological trivial and gapless; for $J_2/J_1>0.32$, it is a $\pi/2$-flux chiral spin liquid, which is topological nontrivial and gapped. These results suggest that there exist two topologically different spin disorder phases in honeycomb $J_1$-$J_2$ Heisenberg model.

\# correspandance author: zou@theory.issp.ac.cn

I.   Introduction

Strong spin frustration and magnetic fluctuations may destroy magnetic long-range order, lead to quantum spin disordered or spin liquid (SL) phase in strongly correlated Mott insulator. As a possible precursor of high-$T_c$ superconductors which might contributes mysterious pairing forces of high-$T_c$ cuprate compounds, quantum spin disordered and SL phases have been attracting much attention in the past three decades [1]. Of these spin frustrated systems, the strongly correlated systems on honeycomb lattice has received great interest in the past decade [3-5], not only due to the Dirac-type linear energy dispersion of the electrons at low energies, but also due to small coordination number z=3 causing large quantum spin fluctuations. Recent studies suggested that the SL or spin disordered phase could emerge from the frustrated spin-1/2 antiferromagnetic (AF) $J_1$-$J_2$ Heisenberg model for intermediate $J_2/J_1$ on honeycomb lattice, though the specific parameter range of which has been greatly debated [5-14].

Currently, the essence of the spin disordered regime of the spin-1/2 AF $J_1$-$J_2$ Heisenberg model on honeycomb lattice is far from reaching agreement. Based on the Schwinger boson mean-field method, Yu *et al*. [6] showed the spin disordered or SL phase lies in 0.21<$J_2/J_1$<0.43; Merno and Ralko [7] suggested that a similar quantum SL regime for 0.2<$J_2/J_1$<0.4. Utilizing the density matrix renormalization group technique, Fisher *et al*. [8] proposed that SL phase lies in 0.2<$J_2/J_1$<0.25, and a plaquette valence bond (PVB) order is possible stable in 0.25<$J_2/J_1$<0.35; while van den Brink *et al*. [9] suggested that the PVB and dimer phases lies in 0.22<$J_2/J_1$<0.35 and $J_2/J_1$>0.35, respectively; and White *et al*. [10] showed that weak PVB and dimer (i.e. staggered VB) phases are stable for 0.26<$J_2/J_1$<0.36 and $J_2/J_1$>0.36, respectively. Within the variational approach, Clark *et al*. [5] gave the SL phase in 0.08<$J_2/J_1$<0.3 and the dimer phase for $J_2/J_1$>0.3; Mezzacapo and Boninsegni [11] suggested the paramagnetic phase in

$0.2<J_2/J_1<0.4$; while Ciolo *et al.* [12] suggested a PVB phase for $0.28<J_2/J_1<0.38$ and a dimer phase for $0.38<J_2/J_1<0.6$; More recently Ferrari *et al.* [13] suggested that from $J_2/J_1=0.23$ to 0.5 the PVB solid and columnar phases which separate at $(J_2/J_1)_c=0.36$ are superior to the SL phase. These show that there exist not only great debates on the parameter range of spin disordered phases, but also much controversial on the nature of spin disordered phases, which appeal for further investigation.

Nevertheless, it is definitely that the SL or spin disordered regime in the parameter range of $0.2<J_2/J_1<0.5$ in the frustrated $J_1$-$J_2$ Heisenberg model on honeycomb lattice, deserves further investigation. In the early time Wen [2] pointed out that SL phases may exhibit different topology, including $\pi$-flux state which is gapless and has trivial topology, or $\pi/2$-flux state, also called chiral SL phase, which has finite gap and nontrivial topology. These spin disordered states can also be classified by symmetry, such as valence bond crystals with broken translational symmetry, and pure SL that have all symmetries restored. In the present frustrated $J_1$-$J_2$ Heisenberg model on honeycomb lattice, no matter what the concrete parameter range of the existence of spin disordered phase is, one may interestedly ask the possible topology of these unusual phases, which motivate us to investigate the zero-temperature topological properties of the present model in the parameter range of $0.2<J_2/J_1<0.5$.

In this paper we investigate in detail the properties of SL of the $J_1$-$J_2$ Heisenberg model in two-dimensional (2D) honeycomb lattice. By introducing local bond parameters [15] and calculating the Chern number of various phases, we obtain a mean-field phase diagram. It is particularly interested when focusing on the well-known spin disorder regime $0.2<J_2/J_1<0.45$, we find two distinct spin disordered phases: one is topological trivial and gapless, the other one is topological nontrivial and gapped; the quantum critical point $(J_2/J_1)_c \approx 0.31$ is also confirmed by the thermodynamic properties. The rest of this paper is organized as follows: in Sec. II we describe the $J_1$-$J_2$ Heisenberg model Hamiltonian and theoretical methods; then in Sec. III we present the zero-temperature ground-state topological properties and the phase diagram; finally, Sec. IV is devoted to summary.

## II.  Model Hamiltonian & Theoretical Methods

We start with the spin-1/2 Heisenberg model with AF $J_1$ and $J_2$ interactions

$$H = J_1 \sum_{\langle i,j \rangle} \vec{S}_i \cdot \vec{S}_j + J_2 \sum_{\langle\langle i,j \rangle\rangle} \vec{S}_i \cdot \vec{S}_j \tag{1}$$

on a 2D honeycomb lattice, here $J_1$ and $J_2$ denotes the nearest neighbor (NN) and next-nearest neighbor (NNN) spin coupling strengths, respectively.

As addressed in the preceding, it has well established that large $J_1$ interaction is in favor of the Neel AF phase and large $J_2$ interaction favors the stripe AF phase. These two competing interactions lead to strong spin frustrations and strong spin fluctuations in the intermediate $J_1$-$J_2$. To discuss the topological characters of the spin disorder phases, similar to Raghu et al. in Ref. [15], we consider local bond order parameters by introducing the Hubbard-Stratonovich fields, including $\chi_{ij}^{\mu} = \sum_{\alpha\beta} c_{i\alpha}^{+} \sigma_{\alpha\beta}^{\mu} c_{j\beta}$ ( $\mu = 0,1,2,3$, $\sigma^{\mu} = (1, \vec{\sigma})$) for NNN $i, j$ and $\kappa_{ij}^{\mu} = \sum_{\alpha\beta} a_{i\alpha}^{+} \sigma_{\alpha\beta}^{\mu} b_{j\beta}$ for NN $i, j$, as seen in Fig. 1. Physically, $\langle \chi_{ij}^{0} \rangle = \langle c_{i\uparrow}^{+} c_{j\uparrow} + c_{i\downarrow}^{+} c_{j\downarrow} \rangle \neq 0$ means the quantum anomalous Hall phase, in which the Chern number $n_C^{\uparrow} + n_C^{\downarrow} \neq 0$, and $\langle \chi_{ij}^{3} \rangle = \langle c_{i\uparrow}^{+} c_{j\uparrow} - c_{i\downarrow}^{+} c_{j\downarrow} \rangle \neq 0$ means the quantum spin Hall phase, in which $n_C^{\uparrow} - n_C^{\downarrow} \neq 0$.

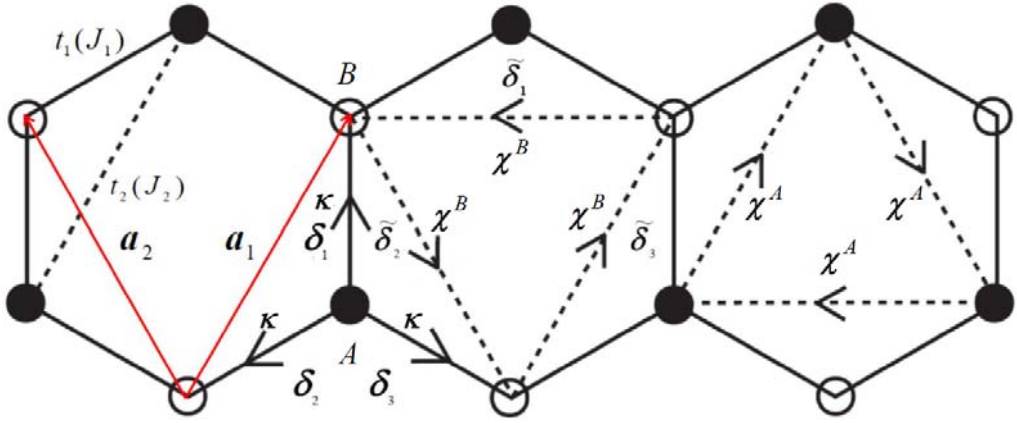

FIG. 1. Schematic diagram of honeycomb lattice and interactions considered in our paper. The bond parameters $\chi_A$, $\chi_B$ are 4-vector associated with the NNN links $\tilde{\delta}_i$ while $\kappa$ associated with the NN links $\delta_i$.

The superexchange spin interactions in Eq.(1) could be recast using the identities

$$J_2 \vec{S}_i \cdot \vec{S}_j = \frac{J_2}{8}\left[-3(\chi_{ij}^0)^\dagger \chi_{ij}^0 + (\chi_{ij}^1)^\dagger \chi_{ij}^1 + (\chi_{ij}^2)^\dagger \chi_{ij}^2 + (\chi_{ij}^3)^\dagger \chi_{ij}^3\right]$$

$$J_1 \vec{S}_i \cdot \vec{S}_j = \frac{J_1}{8}\left[-3(\kappa_{ij}^0)^\dagger \kappa_{ij}^0 + (\kappa_{ij}^1)^\dagger \kappa_{ij}^1 + (\kappa_{ij}^2)^\dagger \kappa_{ij}^2 + (\kappa_{ij}^3)^\dagger \kappa_{ij}^3\right] \quad (2)$$

A translational and rotational invariant ansatz of the bond parameters is chosen as

$$\langle \chi_{Ai,Ai+\tilde{\delta}}^\mu \rangle = \chi_A^\mu, \quad \langle \chi_{Bj,Bj+\tilde{\delta}}^\mu \rangle = \chi_B^\mu, \quad \langle \kappa_{Ai,Ai+\delta}^\mu \rangle = \kappa^\mu, \quad (3)$$

giving rise to a $4\times 4$ Hamiltonian that is readily diagonalized.

From the symmetry consideration, we have $\chi^1 = \chi^2 = \kappa^1 = \kappa^2 = 0$, which implies that there is no hybridization between spin-up and spin-down fermions. In this approximation the effective Hamiltonian could be rewritten as a block matrix

$$H = \sum_k \begin{pmatrix} a_{k\uparrow}^+ & b_{k\uparrow}^+ & a_{k\downarrow}^+ & b_{k\downarrow}^+ \end{pmatrix} \begin{pmatrix} H^\uparrow(\vec{k}) & 0 \\ 0 & H^\downarrow(\vec{k}) \end{pmatrix} \begin{pmatrix} a_{k\uparrow} \\ b_{k\uparrow} \\ a_{k\downarrow} \\ b_{k\downarrow} \end{pmatrix} + E_0, \quad (4)$$

where the diagonal elements are $H^{s}_{11(22)} = \frac{J_2}{8}\sum_{\tilde{\delta}}\left[(-3\chi^{0}_{A(B)} + s\chi^{3}_{A(B)})e^{-ik\cdot\tilde{\delta}} + h.c.\right]$ and the nondiagonal elements are $(H^{s}_{12})^{*} = H^{s}_{21} = \left(-\frac{3J_1}{8}\kappa^0 + s\frac{J_1}{8}\kappa^3\right)\left(\sum_{\delta}e^{-ik\cdot\delta}\right)$, $s = \pm$

represent the spin-up and spin-down branches; and the constant energy part is

$$E_0 = -\frac{3}{8}NJ_1\left(-3|\kappa^0|^2 + |\kappa^3|^2\right) - \frac{3}{8}NJ_2\left(-3|\chi^0_A|^2 - 3|\chi^0_B|^2 + |\chi^3_A|^2 + |\chi^3_B|^2\right).$$

Diagonalizing the mean-field Hamiltonian one obtains the groundstate energy. Minimizing the groundstate energy we get a set of self-consistent equations of the order parameters as follows:

$$S = \frac{1}{2N}\sum_{k}\left[|U^{\uparrow}_{11}|^2 f(\varepsilon_1) + |U^{\uparrow}_{12}|^2 f(\varepsilon_2) - |U^{\downarrow}_{11}|^2 f(\varepsilon_3) - |U^{\downarrow}_{12}|^2 f(\varepsilon_4)\right]$$

$$= -\frac{1}{2N}\sum_{k}\left[|U^{\uparrow}_{21}|^2 f(\varepsilon_1) + |U^{\uparrow}_{22}|^2 f(\varepsilon_2) - |U^{\downarrow}_{21}|^2 f(\varepsilon_3) - |U^{\downarrow}_{22}|^2 f(\varepsilon_4)\right]$$

$$\chi^0_A = \frac{1}{N}\sum_{k}\left[|U^{\uparrow}_{11}|^2 f(\varepsilon_1) + |U^{\uparrow}_{12}|^2 f(\varepsilon_2) + |U^{\downarrow}_{11}|^2 f(\varepsilon_3) + |U^{\downarrow}_{12}|^2 f(\varepsilon_4)\right]e^{ik\cdot\tilde{\delta}}$$

$$\chi^3_A = \frac{1}{N}\sum_{k}\left[|U^{\uparrow}_{11}|^2 f(\varepsilon_1) + |U^{\uparrow}_{12}|^2 f(\varepsilon_2) - |U^{\downarrow}_{11}|^2 f(\varepsilon_3) - |U^{\downarrow}_{12}|^2 f(\varepsilon_4)\right]e^{ik\cdot\tilde{\delta}}$$

$$\chi^0_B = \frac{1}{N}\sum_{k}\left[|U^{\uparrow}_{21}|^2 f(\varepsilon_1) + |U^{\uparrow}_{22}|^2 f(\varepsilon_2) + |U^{\downarrow}_{21}|^2 f(\varepsilon_3) + |U^{\downarrow}_{22}|^2 f(\varepsilon_4)\right]e^{ik\cdot\tilde{\delta}}$$

$$\chi^3_B = \frac{1}{N}\sum_{km}\left[|U^{\uparrow}_{21}|^2 f(\varepsilon_1) + |U^{\uparrow}_{22}|^2 f(\varepsilon_2) - |U^{\downarrow}_{21}|^2 f(\varepsilon_3) - |U^{\downarrow}_{22}|^2 f(\varepsilon_4)\right]e^{ik\cdot\tilde{\delta}}$$

$$\kappa^0 = \frac{1}{N}\sum_{k}\left[U^{\uparrow*}_{11}U^{\uparrow}_{21}f(\varepsilon_1) + U^{\uparrow*}_{12}U^{\uparrow}_{22}f(\varepsilon_2) + U^{\downarrow*}_{11}U^{\downarrow}_{21}f(\varepsilon_3) + U^{\downarrow*}_{12}U^{\downarrow}_{22}f(\varepsilon_4)\right]e^{ik\cdot\delta}$$

$$\kappa^3 = \frac{1}{N}\sum_{k}\left[U^{\uparrow*}_{11}U^{\uparrow}_{21}f(\varepsilon_1) + U^{\uparrow*}_{12}U^{\uparrow}_{22}f(\varepsilon_2) - U^{\downarrow*}_{11}U^{\downarrow}_{21}f(\varepsilon_3) - U^{\downarrow*}_{12}U^{\downarrow}_{22}f(\varepsilon_4)\right]e^{ik\cdot\delta}$$

where S is the spin and U is the transformation matrix for the diagonalization. Numerically solving these equations we could get the evolutions of these order parameters with the variations of $J_2$ and $J_1$.

On the other hand, the block Hamiltonian Eq.(4) could be further expressed as the form of

$$H(\vec{k}) = \begin{pmatrix} \varepsilon^{\uparrow}(\vec{k}) + \vec{d}^{\uparrow}(\vec{k}) \cdot \vec{\sigma} & 0 \\ 0 & \varepsilon^{\downarrow}(\vec{k}) + \vec{d}^{\downarrow}(\vec{k}) \cdot \vec{\sigma} \end{pmatrix}, \qquad (5)$$

where

$$d_x^s = \left(-\frac{J_1}{4}\text{Re}(\kappa^0) + s\frac{J_1}{4}\text{Re}(\kappa^3)\right)\sum_{\delta}\cos(\vec{k}\cdot\delta) + \frac{J_1}{4}\left(-\text{Im}(\kappa^0) + s\,\text{Im}(\kappa^3)\right)\sum_{\delta}\sin(\vec{k}\cdot\delta),$$

$$d_y^s = \left(\frac{J_1}{4}\text{Re}(\kappa^0) + s\frac{J_1}{4}\text{Re}(\kappa^3)\right)\sum_{\delta}\sin(\vec{k}\cdot\delta) - \frac{J_1}{4}\left(\text{Im}(\kappa^0) + s\,\text{Im}(\kappa^3)\right)\sum_{\delta}\cos(\vec{k}\cdot\delta),$$

$$d_z^s = \frac{J_2}{4}\sum_{\tilde{\delta}}\begin{bmatrix}\text{Re}(-\chi_A^0 + s\chi_A^3 + \chi_B^0 - s\chi_B^3)\cos(\vec{k}\cdot\tilde{\delta}) \\ +\text{Im}(-\chi_A^0 + s\chi_A^3 + \chi_B^0 - s\chi_B^3)\sin(\vec{k}\cdot\tilde{\delta})\end{bmatrix}.$$

For the present system, the Berry curvature is given by [16]

$$\Omega_{k_x,k_y} = \frac{1}{2}\hat{d}\cdot\partial_x\hat{d}\times\partial_y\hat{d} = \frac{1}{2}\varepsilon^{\alpha\beta\gamma}\frac{\partial\hat{d}_\alpha(\vec{k})}{\partial k_x}\frac{\partial\hat{d}_\beta(\vec{k})}{\partial k_y}\hat{d}_\gamma(\vec{k}) \qquad (6)$$

and the Chern number is expressed as

$$n_{chern} = \frac{2\pi}{S}\sum_k \Omega_{k_x,k_y}[n_+(\vec{k}) - n_-(\vec{k})] \qquad (7)$$

where the renormalized vector $\hat{d}_\alpha(\vec{k}) = d_\alpha(\vec{k})/d(\vec{k})$ with $d(\vec{k}) = \sqrt{d_x^2 + d_y^2 + d_z^2}$. From which we could numerically evaluate the Chern number of the honeycomb $J_1$-$J_2$ Heisenberg model. We focus our attention on the well-known spin disordered regime, so as to uncover the topology of the spin disordered phases. We adopt the earlier results on the NAF and SAF phases in Ref.[6], since the long-range AFM order parameters of these two phases could not be described by the present bond order parameter $\kappa^\mu$ and $\chi^\mu$.

### III. Zero-Temperature Topology & Phase Diagram

To have a whole understanding for the spin disordered phase, we further study the topological properties of the honeycomb $J_1$-$J_2$ Heisenberg system. We present the dependence of local order parameters $\chi^\mu$ and $\kappa^\mu$ on $J_1/J_2$ in Fig. 2. Combining the proceeding results, as well as those of Yu et al. [6] and many earlier literature, the phase diagram consists of four different areas: Neel AF, spin disordered phases with real and with complex order parameters, and striped AF, corresponding to different parameter range of $J_1/J_2$. The local order parameters divide the spin disorder area into two parts obviously: $\chi^0 = 0$ for $0.21 \leq J_2/J_1 \leq 0.32$, and $\chi_A^0 = -\chi_B^0$ are pure imaginaries for $0.32 < J_2/J_1 \leq 0.43$. $\kappa^0$ is real in the whole parameter range of spin disorder phase $0.21 \leq J_2/J_1 \leq 0.43$. This shows that there exist two distinct spin disordered phases in the $J_1 - J_2$ Heisenberg systems on honeycomb lattice.

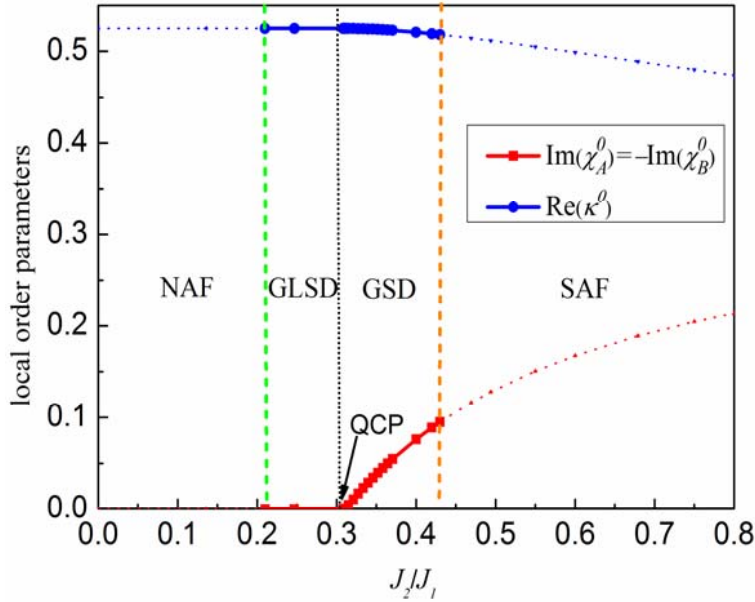

FIG. 2. Mean-field phase diagram as a function of $J_2/J_1$. NAF, GLSD, GSD and SAF denote the Neel AF phase, gapless and gapped spin disorder phases, and striped AF phase, respectively. The black dotted line denotes the position of the quantum critical point.

To further understand the two distinct spin disordered phases, we calculate the corresponding dispersion relations of spin excitations. The numerical results are shown in Fig. 3. For $0.21 \leq J_2/J_1 \leq 0.32$, the lower spectrum and upper spectrum intersect at the $K$ point, forming a Dirac cone structure similar to graphene. For $0.32 \leq J_2/J_1 \leq 0.43$, the dispersion relation of spin excitations opens a gap at the $K$ point. The energy gap of the dispersion relations as a function of $J_2/J_1$ is shown in Fig. 3. The gapless and gapped properties of spin excitation further distinguish these two spin disordered phases.

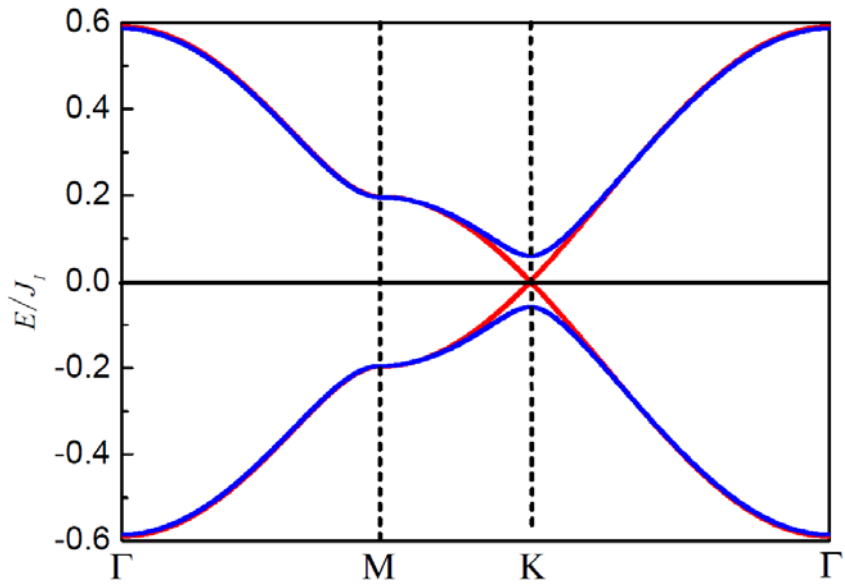

FIG. 3. The dispersion relations of spin excitations for $J_2/J_1 = 0.25$ (the red line) and for $J_2/J_1 = 0.4$ (the blue line).

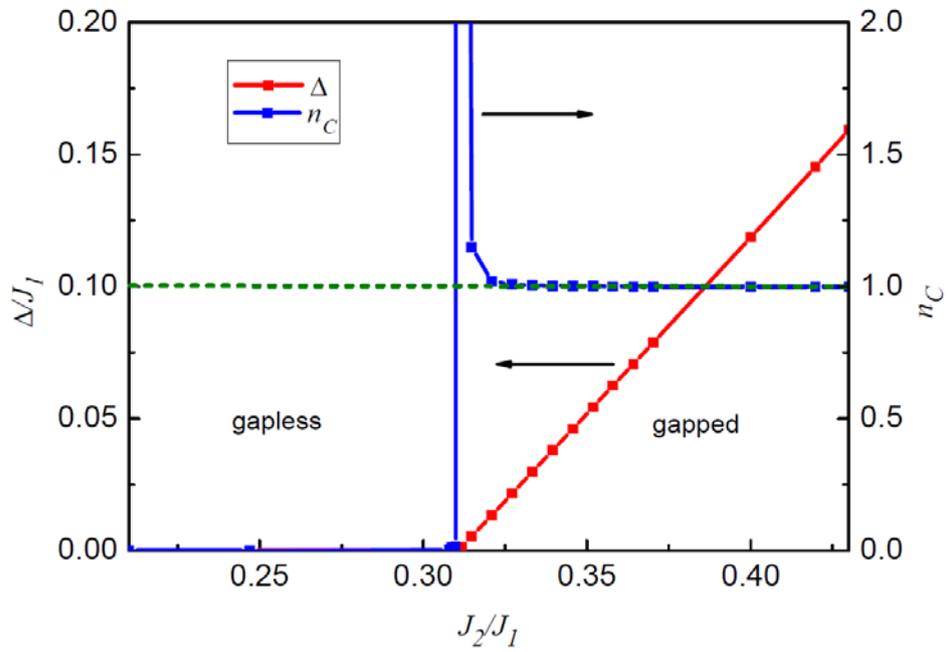

FIG. 4. $J_2/J_1$--dependence of the energy gap $\Delta$ (the red line) and Chern number $n_C$ (the blue line).

In order to get insight the topological properties of two spin disordered phases, we calculate the Chern numbers of the system in these two parameter ranges. Chern number $n_C^\uparrow = n_C^\downarrow$ as a function of $J_2/J_1$ is shown in Fig. 4. One finds that a transition occurs at $J_2/J_1 = 0.32$. For $J_2/J_1 \leq 0.32$, $n_C^\uparrow = n_C^\downarrow = 0$, the system has trivial topology. For $J_2/J_1 \geq 0.32$, $n_C^\uparrow = n_C^\downarrow = 1$, the system is in a quantum anomalous Hall state with nontrivial topology. The topological quantum number has a divergence at the phase transition point $J_2/J_1 = 0.32$, as seen in Fig. 4.

Physically, our system with large $J_2$ is similar to the Haldane model [16], while the pure imaginary $\chi^0$ produces a $\pi/2$-flux in the corresponding loop path, playing the role of an effective external magnetic field, and the real $\kappa^0$ plays the role of hopping $t$. As a consequence, the gapless and gapped spin disordered states are two topologically distinguishable phases.

**IV. Concluding Remarks**

The microscopic origin of such unusual topological properties mainly arises from the spin frustrations and spin fluctuations of the NNN spin coupling $J_2$, which behaves similar properties as the flux term in the Haldane model [16], a large $J_2$ term drives quasiparticles to acquire a topological phase similar to the spin-orbital coupling in the Kane-Mele model [17]. Also, as pointed in the early studies [5-14], the first spin disorder phase corresponds to isotropic plaquette valence bond solid, and the second one to anisotropic dimer or columnar valence bond solid. One naturally expects that the former is gapless, and the latter is gapped, since the spatial sixfold rotation symmetry is broken in the latter.

In summary, we have shown different topological properties of the spin disordered phases of the $J_1$-$J_2$ Heisenberg model on honeycomb lattice, i.e., the first spin disordered phase in $J_2/J_1<0.32$ is gapless and trivial topology, and the second one in $J_2/J_1>0.32$ is topological nontrivial chiral phase with finite gap. These results clearly demonstrate that the there exist two topologically distinct spin disorder/liquid phases in the antiferromagnetic $J_1$-$J_2$ Heisenberg system on honeycomb lattice.

**Acknowledgement**

This work is supported by the National Natural Science Foundation of China under Grant Nos. 11774350, 11474287, 11534010, and 11734002. Numerical calculations were performed at the Center for Computational Science of CASHIPS and Tianhe II of CSRC.